\documentclass[useAMS,usenatbib]{mn2e}
\usepackage{graphicx}
\usepackage{lscape}

\newcommand{\Msun}{\mbox{M$_{\odot}$}}
\newcommand{\Lsun}{\mbox{$L_{\odot}$}} 
\newcommand{\Rsun}{\mbox{$R_{\odot}$}} 
\newcommand{\ltsimeq}{\raisebox{-0.6ex}{$\,\stackrel
        {\raisebox{-.2ex}{$\textstyle <$}}{\sim}\,$}}

\title[An M4-T8.5 binary]{The discovery of an M4+T8.5 binary system}
\author[Ben Burningham et al]{Ben Burningham$^{1}$\thanks{E-mail:
    B.Burningham@herts.ac.uk},D.J. Pinfield$^{1}$,
  S. K. Leggett$^{2}$,C. G. Tinney$^{3}$, M.C.Liu$^{4}$\thanks{Alfred P. Sloan Research Fellow},D. Homeier$^5$,
\newauthor
 A.A. West$^{6}$, A. Day-Jones$^{1}$, N. Huelamo$^{7}$,
 T.J.Dupuy$^{3}$, Z.Zhang$^{1}$,D.N.Murray$^{1}$,
\newauthor
N. Lodieu$^{8}$, D. Barrado y Navascu\'es$^{7}$, S. Folkes$^1$, M.C.Galvez-Ortiz$^{1}$,  H. R. A. Jones$^{1}$,
\newauthor
 P. W. Lucas$^{1}$, M. Morales Calderon$^{7}$, M. Tamura$^{9}$\\
$^{1}$Centre for Astrophysics Research, Science and Technology
Research Institute, University of Hertfordshire, Hatfield AL10 9AB \\
$^{2}$Gemini Observatory, 670 N. A'ohoku Place, Hilo, HI 96720, USA \\
$^{3}$University of New South Wales, Australia\\
$^{4}$Institute for Astronomy, University of Hawai`i, 2680 Woodlawn
Drive, Honolulu, HI 96822 \\
$^{5}$Institut fur Astrophysik, Georg-August-Universitat,
Friedrich-Hund-Platz 1, 37077 Gottingen, Germany \\
$^{6}$MIT Kavli Institute for Astrophysics and Space Research, 77
    Massachusetts Avenue, Cambridge, MA 02139 \\
$^{7}$Laboratorio de Astrof\'isica Espacial y F\'isica
     Fundamental,INTA, P.O. Box 78, E--28691 Villanueva de la Canada (Madrid), Spain \\
$^{8}$Instituto de Astrof\'isica de Canarias, 38200 La Laguna, Spain \\
$^{9}$ National Astronomical Observatory, Mitaka, Tokyo 181-8588\\
}

\begin{document}
%
%  These Macros are taken from the AAS TeX macro package version 4.0.
%  Include this file in your LaTeX source only if you are not using
%  the AAS TeX macro package and need to resolve the macro definitions
%  in the BibTeX entries returned by the ADS abstract service.
%
%  If you plan not to use this file to resolve the journal macros
%  rather than the whole AAS TeX macro package, you should save the
%  file as ``aas_macros.sty'' and then include it in your paper by
%  using a construct such as:
%	\documentstyle[11pt,aas_macros]{article}
%
%  For more information on the AASTeX macro package, please see the URL
%	http://www.aas.org/publications/aastex.html
%  For more information about ADS abstract server, please see the URL
%	http://adswww.harvard.edu/ads_abstracts.html
%

% Abbreviations for journals.  The object here is to provide authors
% with convenient shorthands for the most "popular" (often-cited)
% journals; the author can use these markup tags without being concerned
% about the exact form of the journal abbreviation, or its formatting.
% It is up to the keeper of the macros to make sure the macros expand
% to the proper text.  If macro package writers agree to all use the
% same TeX command name, authors only have to remember one thing, and
% the style file will take care of editorial preferences.  This also
% applies when a single journal decides to revamp its abbreviating
% scheme, as happened with the ApJ (Abt 1991).

\def\aj{\rm{AJ}}                   % Astronomical Journal
\def\araa{\rm{ARA\&A}}             % Annual Review of Astron and Astrophys
\def\apj{\rm{ApJ}}                 % Astrophysical Journal
\def\apjl{\rm{ApJ}}                % Astrophysical Journal, Letters
\def\apjs{\rm{ApJS}}               % Astrophysical Journal, Supplement
\def\ao{\rm{Appl.~Opt.}}           % Applied Optics
\def\apss{\rm{Ap\&SS}}             % Astrophysics and Space Science
\def\aap{\rm{A\&A}}                % Astronomy and Astrophysics
\def\aapr{\rm{A\&A~Rev.}}          % Astronomy and Astrophysics Reviews
\def\aaps{\rm{A\&AS}}              % Astronomy and Astrophysics, Supplement
\def\azh{\rm{AZh}}                 % Astronomicheskii Zhurnal
\def\baas{\rm{BAAS}}               % Bulletin of the AAS
\def\jrasc{\rm{JRASC}}             % Journal of the RAS of Canada
\def\memras{\rm{MmRAS}}            % Memoirs of the RAS
\def\mnras{\rm{MNRAS}}             % Monthly Notices of the RAS
\def\pra{\rm{Phys.~Rev.~A}}        % Physical Review A: General Physics
\def\prb{\rm{Phys.~Rev.~B}}        % Physical Review B: Solid State
\def\prc{\rm{Phys.~Rev.~C}}        % Physical Review C
\def\prd{\rm{Phys.~Rev.~D}}        % Physical Review D
\def\pre{\rm{Phys.~Rev.~E}}        % Physical Review E
\def\prl{\rm{Phys.~Rev.~Lett.}}    % Physical Review Letters
\def\pasp{\rm{PASP}}               % Publications of the ASP
\def\pasj{\rm{PASJ}}               % Publications of the ASJ
\def\qjras{\rm{QJRAS}}             % Quarterly Journal of the RAS
\def\skytel{\rm{S\&T}}             % Sky and Telescope
\def\solphys{\rm{Sol.~Phys.}}      % Solar Physics
\def\sovast{\rm{Soviet~Ast.}}      % Soviet Astronomy
\def\ssr{\rm{Space~Sci.~Rev.}}     % Space Science Reviews
\def\zap{\rm{ZAp}}                 % Zeitschrift fuer Astrophysik
\def\nat{\rm{Nature}}              % Nature
\def\iaucirc{\rm{IAU~Circ.}}       % IAU Cirulars
\def\aplett{\rm{Astrophys.~Lett.}} % Astrophysics Letters
\def\apspr{\rm{Astrophys.~Space~Phys.~Res.}}
                % Astrophysics Space Physics Research
\def\bain{\rm{Bull.~Astron.~Inst.~Netherlands}} 
                % Bulletin Astronomical Institute of the Netherlands
\def\fcp{\rm{Fund.~Cosmic~Phys.}}  % Fundamental Cosmic Physics
\def\gca{\rm{Geochim.~Cosmochim.~Acta}}   % Geochimica Cosmochimica Acta
\def\grl{\rm{Geophys.~Res.~Lett.}} % Geophysics Research Letters
\def\jcp{\rm{J.~Chem.~Phys.}}      % Journal of Chemical Physics
\def\jgr{\rm{J.~Geophys.~Res.}}    % Journal of Geophysics Research
\def\jqsrt{\rm{J.~Quant.~Spec.~Radiat.~Transf.}}
                % Journal of Quantitiative Spectroscopy and Radiative Transfer
\def\memsai{\rm{Mem.~Soc.~Astron.~Italiana}}
                % Mem. Societa Astronomica Italiana
\def\nphysa{\rm{Nucl.~Phys.~A}}   % Nuclear Physics A
\def\physrep{\rm{Phys.~Rep.}}   % Physics Reports
\def\physscr{\rm{Phys.~Scr}}   % Physica Scripta
\def\planss{\rm{Planet.~Space~Sci.}}   % Planetary Space Science
\def\procspie{\rm{Proc.~SPIE}}   % Proceedings of the SPIE

\let\astap=\aap
\let\apjlett=\apjl
\let\apjsupp=\apjs
\let\applopt=\ao

\maketitle

\begin{abstract}
We report the discovery of a T8.5 dwarf, which is a companion to
the M4 dwarf Wolf~940. 
At a distance of 12.50$^{+0.75}_{-0.67}$~pc, the angular separation of
32\arcsec\ corresponds to a projected separation of 400 AU.
The M4 primary displays no H$\alpha$ emission, and we apply the
age-activity relations of West et al. to place a lower limit on the age
of the system of 3.5 Gyr. Weak H$\alpha$ absorption suggests some
residual activity and we estimate an upper age limit of 6~Gyr.
We apply the relations of Bonfils et al for $V-K_s$ and $M_{K_s}$
to determine the metallicity, ${\rm [Fe/H]} = -0.06 \pm 0.20$ for
Wolf~940A, and by extension the T8.5 secondary, Wolf 940B. 
We have obtained $JHK$ NIRI spectroscopy and $JHKL'$ photometry of
Wolf 940B, and use these data, in combination with theoretical
extensions, to determine its bolometric flux, $F_{bol} = 1.75 \pm 0.18
\times 10^{-16} {\rm W}m^{-2}$ and thus its luminosity $\log
(L_{*}/\Lsun) = -6.07 \pm 0.04$.
Using the age constraints for the system, and evolutionary structural
models of Baraffe et al. we determine $T_{\rm eff} = 570 \pm 25$K and
$\log g = 4.75-5.00$ for Wolf~940B, based on its bolometric
luminosity. 
This represents the first
determination of these properties for a T8+ dwarf that does not rely on the
fitting of T-dwarf spectral models.
This object represents the first system containing a
T8+ dwarf for which fiducial constraints on its properties are
available, and we compare its spectra with those
of the latest very cool BT-Settl models. 
This clearly demonstrates that the use of the ($W_J$,$K/J$) spectral
ratios (used previously to constrain $T_{\rm eff}$ and $\log g$) would
have over-estimated $T_{\rm eff}$ by $\sim 100$K.
\end{abstract}


\section{Introduction}
\label{sec:intro}

The advent of the most recent generation of large imaging surveys
\citep[e.g. ][ and soon the VISTA surveys]{ukidss,cfbds} has
facilitated the identification of brown dwarfs with later spectral types
than the latest T~dwarfs found  using 2MASS, DENIS or SDSS.
For example, the UKIRT Infrared Deep Sky Survey
(UKIDSS) Large Area Survey \citep[LAS; see][]{ukidss}, which as of
Data Release 4 (DR4) probes nearly 3 times the searchable volume of 2MASS for
such objects,
contains at least four T~dwarfs with spectral types later than T8 (in
addition to that identified in this work),
which have recently been classified by \citet{ben08}: the T9 dwarfs
ULAS~J003402.77-005206.7 (hereafter ULAS~0034),
CFBDS~J005910.90-011401.3 (CFBDS~0059)and
ULAS~J133553.45+113005.2 (ULAS~1335); and the T8.5 dwarf
ULAS~J123828.51+095351.3 (ULAS~1238) \citep{warren07,delorme08,ben08}.
The physical properties of these objects have been estimated by
fitting various model spectra to near- and mid-infrared data (where available).
Effective temperature ($T_{\rm eff}$) estimates vary from as cool as
500--550~K for ULAS~1335 \citep{leggett09} to as warm as
600--650~K for ULAS~0034 \citep{warren07}. 
Since parallax determinations are not yet available for these objects,
these estimates are somewhat uncertain, based as they are on early
generations of atmospheric models that are still under development.
Also, current indications are that surface gravity and metallicity are largely
degenerate as far as near-infrared spectral fitting is concerned,
which adds another layer of uncertainty to the parameters derived for
such objects to date.

Since the low-temperature extreme of the brown dwarf regime is of
particular interest for determining the form of the substellar initial
mass function \citep[e.g.][]{burgasser04}, it is extremely desirable
that atmospheric models in this regime are robustly constrained. 
Furthermore, the sub-600~K temperature regime overlaps with the
warm-exoplanet regime, and such cool brown dwarfs provide excellent
laboratories for improving the substellar atmospheric models which
will be key to interpreting observations over the coming years.
The discovery of T8+ dwarfs in binary systems with stellar primaries
is of central importance for improving the current generation of
atmospheric models, since we can use the properties of the primary
star as fiducial constraints on the properties of the substellar secondary
\citep[e.g.][]{pinfield06, burgasser05}.

The term ``benchmark'' is broadly applied to objects for
which at least some properties may be determined with minimal
reference to models (although the degree of
reference to models that is required to determine their properties
varies). 
With the exception of a few T~dwarfs whose ages and metallicities may
be gleaned from studies of the young clusters or moving
groups of which they are members \citep[e.g. Hyades - ][]{bouvier08},
the majority of T~dwarf benchmarks have been found in binary systems.
Indeed, one of the first unequivocally confirmed brown dwarfs, Gl229B
\citep{nakajima1995}, was found as a companion of an early type
M~dwarf, studies of which have yielded improved constraints on the
system properties \citep{leggett02}.
Other notable T~dwarfs in such systems include HN~PegB
(T2.5$\pm$0.5) and HD~3651B (T7.5$\pm$0.5) \citep{mugrauer06,luhman07}, which are
companions to well studied main sequence stars (G0V and K0V,
respectively).

The multiple systems Gl~570 \citep[K4V, M1.5V, M3V \& T7.5;
][]{burgasser2000,geballe01} and $\epsilon$ Indi \citep[K4.5V, T1 \&
  T6; ][]{scholz03,markMcG04} also have well constrained ages from studies of
the K and M dwarf members of the systems.
The $\epsilon$ Indi Ba,Bb system offers the prospect of dynamical mass
estimates for the T~dwarf components through direct observation of the
orbital motion within its short, $\sim$15 year, period \citep{markMcG04}.

\citet{liu08} have suggested that brown dwarf binaries with dynamical
mass determinations can serve as benchmark systems with comparable, or
better, constraints on the brown dwarf gravities than in the case of
wide companions of known age. 
They have demonstrated this approach with the first mass determination for a T
 dwarf binary, the T5+T5.5 system 2MASS~1534-2952AB, though the long
 orbital periods ($>10$ years) mean that some patience will be needed
 until a larger sample of this type of benchmark is available.

We report here on the discovery of a T8+ object,
identified as a low-mass companion to the M4 dwarf Wolf~940, and
explore its potential use as a benchmark object.

\section{A new T8+ dwarf}
\label{sec:ident}

\begin{figure}
\begin{center}
\includegraphics[height=200pt, angle=0]{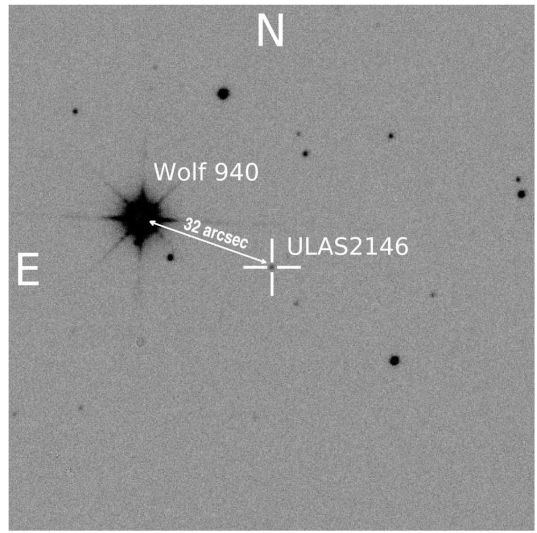}
\caption{A 2'$\times$2' $J$-band finding chart for ULAS~2146 taken from
the UKIDSS database.}
\label{fig:finder}
\end{center}
\end{figure}

Our searches of the UKIRT Infrared Deep Sky Survey (UKIDSS) Large Area
Survey \citep[LAS; see][]{ukidss} have been successful at identifying
late-type T~dwarfs
\citep[e.g.][]{lod07,warren07,pinfield08,ben08}. 
Using the same search methodology as previously described in detail in
\citet{pinfield08}, we identified ULAS~J214638.83-001038.7 (hereafter
ULAS~2146) as a candidate very 
late-T~dwarf, with $YJH$ colours reminiscent of other T8+ dwarfs (it
was undetected in $K$).
The source was observed as part of the LAS in $YJHK$ for 40 seconds in
each filter \citep[see ][]{ukidss}, and the results of these
observations are summarised in Table~\ref{tab:photobs}. 
The subsequent photometric and spectroscopic follow-up, which resulted in its
classification as a T8.5 dwarf, are described in the following sub-sections.
Figure~\ref{fig:finder} shows a UKIDSS $J$-band finding chart for this
object. 

\subsection{Near-infrared photometry}
\label{subsec:photo}

Near-infrared follow-up photometry was obtained using the UKIRT Fast
Track Imager \citep[UFTI;][]{roche03} mounted on UKIRT , and the
Long-slit Infrared Imaging Spectrograph \citep[LIRIS;][]{Manchado98}
mounted on the William Herschel Telescope on La Palma.
Image mosaics were produced using sets of jittered images, with
individual exposure times, jitter patterns and number of repeats given
in Table~\ref{tab:photobs}.
The data were dark subtracted, flatfield corrected, sky subtracted and
mosaiced using ORAC-DR for the UFTI data, and LIRIS-DR for the LIRIS data.

We calibrated our UFTI observations using
UKIRT Faint Standards \citep{ukirtfs}, with a standard observed at a
similar airmass for each target. 
All UFTI data were obtained under photometric conditions, with seeing
better than 0.9\arcsec. 
Photometry was performed using apertures with radii approximately
$\sqrt 3$ times the seeing, which was stable between standard star and
target frames to within less than 0.5 pixels (0.045 \arcsec).

The wider field LIRIS data were obtained in a mixture of photometric
conditions with stable seeing, and thin cirrus in variable seeing (0.8
- 1.2\arcsec). 
Absolute photometry for ULAS~2146, and for a number of fiducial stars,
was obtained during photometric conditions and calibrated using
a UKIRT Faint Standard star (FS29) observed at a similar airmass.   
The zero points for observations obtained in non-photometric conditions
were then determined using these fiducial stars.

We used the spectra of T dwarf spectral standards for types
T2-T9 to synthesise a transformation
between the LIRIS $Y$-band filter and the MKO $Y$-band filter as a
function of spectral type (ST).
We find that :

$$Y_{\sc MKO} = Y_{\sc LIRIS} - (0.022 \times {\rm ST}) - 0.089$$

with a scatter of $\pm 0.01$ (where ${\rm ST} = 2, 3, 4$~etc. for
T2,T3,T4 etc.).   
For earlier type stars (e.g. standards with $Y-J \sim 0$) we find that
the transformation between the two filters is negligible. 
All $Y$-band magnitudes presented here were either measured in, or
transformed into, $Y_{MKO}$.
LIRIS uses a $K_s$ filter, and we transformed the standard
star's $K$ magnitude to $K_s$ using the relations of \citet{carpenter01}. 
The $K_s$ magnitude for ULAS~2146 was transformed to the MKO
system using the transform derived by \citet{pinfield08}.
In both the case of the $Y$- and the $K$-band, transforms were applied
using a spectral type of T8.5 for ULAS~2146 (see
Section~\ref{subsec:spectra}). 

We obtained multiple
observations of ULAS~2146 in several bands to assess any level of
variability.
Table~\ref{tab:photobs} summarises our photometry for ULAS~2146. 
It can be seen that the $J$-band is stable to 5\% over timescales of
up to a year, with the exception of the UKIDSS survey magnitude (which
is $\sim$15\% brighter). We do not consider that the $\sim 2 \sigma$
discrepancies between the follow-up data and the UKIDSS/WFCAM data are
significant.
The $K$-band follow-up data, however, do not agree well, and could reflect some
underlying variability. 
However, the latter measurement derives from a {\sc LIRIS}
observation, and although the conversion to {\sc MKO}  has been well
characterised for earlier spectral types, this may not be the case for
T8+ dwarfs. As such, we defer any detailed discussion of this
discrepancy until multiple measurements on the same system have been obtained.
Overall, we consider that our observations are consistent with a
source that is stable at the $\ltsimeq 5$\% level. 

\subsection{Optical photometry}
\label{subsec:optphot}

We have also obtained optical $z$-band photometry using the ESO Faint Object Spectrograph and Camera (EFOSC2) mounted on the New Technology Telescope
at La Silla, Chile under program 082.C-0399.
These observations are summarised in Table~\ref{tab:photobs}.  
For this optical follow-up we used a gunn $z$-band filter (ESO Z\#623). 
The data were reduced using standard IRAF packages, and then multiple
images of the target were aligned and stacked to increase
signal-to-noise.

We synthesised Sloan~$i'(AB)$, $z'(AB)$ and EFOSC2 Gunn~$z$
photometry for stars with spectral types between B1V and M4V using
spectra drawn from \citet{gunn83}. The resulting
synthetic colours were then used to derive the transform: 

$$z_{\sc EFOSC2}(AB) = z'(AB) - 0.08(i'(AB) - z'(AB)$$ 
 
This allowed the
zero-point in our images to be determined by using SDSS stars as
secondary calibrators.
The uncertainty we quote for our $z$-band photometry incorporates a
scatter of $\sim \pm 0.05$ in the determined zero-points. 
The results of our ground-based follow-up photometry are
given in Table~\ref{tab:photobs}, which also gives the original WFCAM survey
photometry. In all cases we take the measurement with the lowest
uncertainty as our ``final'' value for use elsewhere in the paper.

 \begin{table*}
\begin{tabular}{| c | c c c c c c|}
  \hline
Filter & Magnitude & Instrument & UT Date & Total integration time & $t_{\rm int}$ breakdown  & Photometric? \\
\hline
$z_{EFOSC2}$ & $22.15 \pm 0.13$ & EFOSC2 & 2008 Oct 08 & 3600s & (j =
1,r = 6, $t_{\rm exp}$ = 600s) & n \\
&&&&&&\\
$Y$ & $19.02 \pm 0.08$ & WFCAM & 2007 Oct 12 & 40s & (j = 2, r = 1, $t_{\rm exp}$ = 40s)  & y \\
$Y$ & $18.97 \pm 0.03$ & LIRIS & 2008 Sep 15 & 1000s & (j = 5, r = 5, $t_{\rm exp}$ = 40s) & y \\
&&&&&&\\
$J$ & $18.02 \pm 0.06$ & WFCAM & 2007 Oct 12 & 40s & (m = 4, j = 2, r = 1,  $t_{\rm exp}$ = 5s & y \\
$J$ & $18.21 \pm 0.03$ & UFTI & 2008 Jul 01 & 300s & (j = 5, r = 1, $t_{\rm exp}$ = 60s) & y \\
$J$ & $18.16 \pm 0.02$ & LIRIS & 2008 Sep 15 & 600s & (j = 5, r = 3, $t_{\rm exp}$ = 40s) & y \\
$J$ & $18.16 \pm 0.02$ & LIRIS & 2008 Sep 17 & 1200s & (j = 5, r = 6, $t_{\rm exp}$ = 40s) & n \\
&&&&&&\\ 
$H$ & $18.38 \pm 0.20$ & WFCAM & 2007 Oct 06 & 40s & (j = 4, r = 1, $t_{\rm exp}$ = 10s) & y \\
$H$ & $18.77 \pm 0.03$ & UFTI & 2008 Jul 01 & 900s & (j = 5, r = 3, $t_{\rm exp}$ = 60s) & y \\
&&&&&&\\ 
$K$ & $18.85 \pm 0.05$ & UFTI & 2008 Jul 24 & 900s & (j = 9, r = 2, $t_{\rm exp}$ = 60s) & y \\
$K$ & $19.08 \pm 0.06$ & LIRIS & 2008 Sep 15 & 1800s & (j = 5, r = 18, $t_{\rm exp}$ = 20s) & y \\
&&&&&&\\ 
$L'$ & $15.38 \pm 0.11$ & NIRI & 2008 Oct 20 &  1730s & (j = 4, r = 22.5, $t_{\rm exp}$ = 24x0.8s & y \\
\hline
\end{tabular}
\caption{Summary of the near infrared
  photometric follow-up. The breakdown of each integration
  is given in the final column with the following notation: m = number
  of microsteps; j = number
  of jitter points; r = number of repeats for jitter pattern;
  $t_{\rm exp}$ = exposure time at each jitter point. 
\label{tab:photobs}}

\end{table*}

\subsection{L'-band photometry}
\label{subsec:lband}

$L'$-band imaging of ULAS~2146 was obtained using the Near InfraRed Imager
and Spectrometer \citep[NIRI;][]{hodapp03} on the Gemini North
Telescope on Mauna Kea under program GN-2008B-Q-29 on the night of the
20$^{th}$ October 2008 under photometric conditions.
Individual images were made up of 24 co-added 0.8 second exposures,
which were repeated over a four point offset pattern. 
In total 90 images were recorded for ULAS~2146, with a further 8 images
obtained of the faint standard HD201941.
Each image had its temporally closest neighbour subtracted from it to
remove the rapidly varying and structured sky background.
The sky subtracted images were then flatfielded using a flatfield
frame constructed by median stacking the entire set of images of
ULAS~2146.
The resulting images were offset to the position of the first
image, and median combined to produce the final image.
These observations, and the resulting $L'$-band magnitude, are
summarised in Table~\ref{tab:photobs}.

\subsection{Near-infrared spectroscopy}
\label{subsec:spectra}

Spectroscopy in the $JHK$-bands was obtained for ULAS~2146 using
NIRI on the
Gemini North Telescope on Mauna Kea (under program GN-2008B-Q-29).
All observations were made up of a set of sub-exposures in an ABBA
jitter pattern to facilitate effective background subtraction, with a
slit width of 1\arcsec. 
The length of the A-B jitter was 10\arcsec.
 The observations are summarised in Table~\ref{tab:specobs}.

The NIRI observations were reduced using standard IRAF
Gemini packages. 
A comparison argon arc frame was
used to obtain a dispersion solution, which was applied to the
pixel coordinates in the dispersion direction on the images.
The resulting wavelength-calibrated subtracted AB-pairs had a low-level
of residual sky emission removed by fitting and subtracting this
emission with a set of polynomial functions fit to each pixel row
perpendicular to the dispersion direction, and considering pixel data
on either side of the target spectrum only. 
The spectra were then extracted using a linear aperture, and cosmic
rays and bad pixels removed using a sigma-clipping algorithm.

Telluric correction was achieved by dividing the extracted target
spectra by that of the F4V star HIP103801, observed just before the
target, in the case of the $J$- and $H$-band spectra, whilst for the
$K$-band the A0V star HIP~112179 was used. 
Prior to division, hydrogen lines were removed from the standard star
spectrum by interpolating the stellar
continuum.
Relative flux calibration was then achieved by multiplying through by a
blackbody spectrum with $T_{\rm eff} = 6700$K  for the
F4V standard, and 10,400~K for the A0V standard.
Data obtained for the same spectral regions on different nights were
co-added after relative flux calibration, each weighted by their exposure time.

 \begin{table*}
\begin{tabular}{| c | c c c c c |}
  \hline
Object & UT Date & Integration time & Instrument &
Spectral region \\
\hline
  ULAS~2146  & 2008 Aug 18 & 12x300s & NIRI & $J$ \\
            & 2008 Aug 21 & 12x300s & NIRI & $H$ \\
            & 2008 Aug 23 & 16x224s & NIRI & $K$ \\

\hline
\end{tabular}
\caption{Summary of the near-infrared spectroscopic observations.
\label{tab:specobs}}

\end{table*}

\begin{figure}
\includegraphics[height=250pt, angle=90]{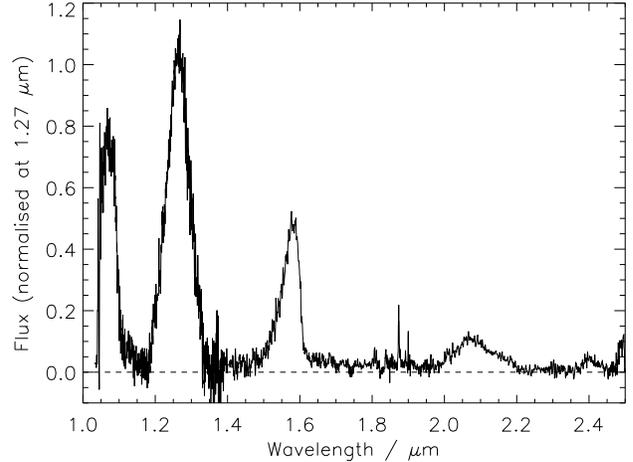}
\caption{The NIRI $JHK$ spectrum for ULAS~2146.}
\label{fig:spec1}
\end{figure}

The spectra were then normalised
using the measured near-infrared photometry to place the spectra on an
absolute flux scale.
The UFTI $JHK$ photometry was used for this purpose since all three
bands were obtained on the same instrument, with MKO filters, within
the shortest interval available in our data for all three bands (see
Table~\ref{tab:photobs}).
The normalised spectrum was rebinned by a factor of three to increase the
signal-to-noise, whilst avoiding under-sampling of the spectral resolution.  
The resultant $JHK$ spectrum for ULAS~2146 is shown in
Figure~\ref{fig:spec1}.

\begin{figure*}
\includegraphics[height=300pt, angle=0]{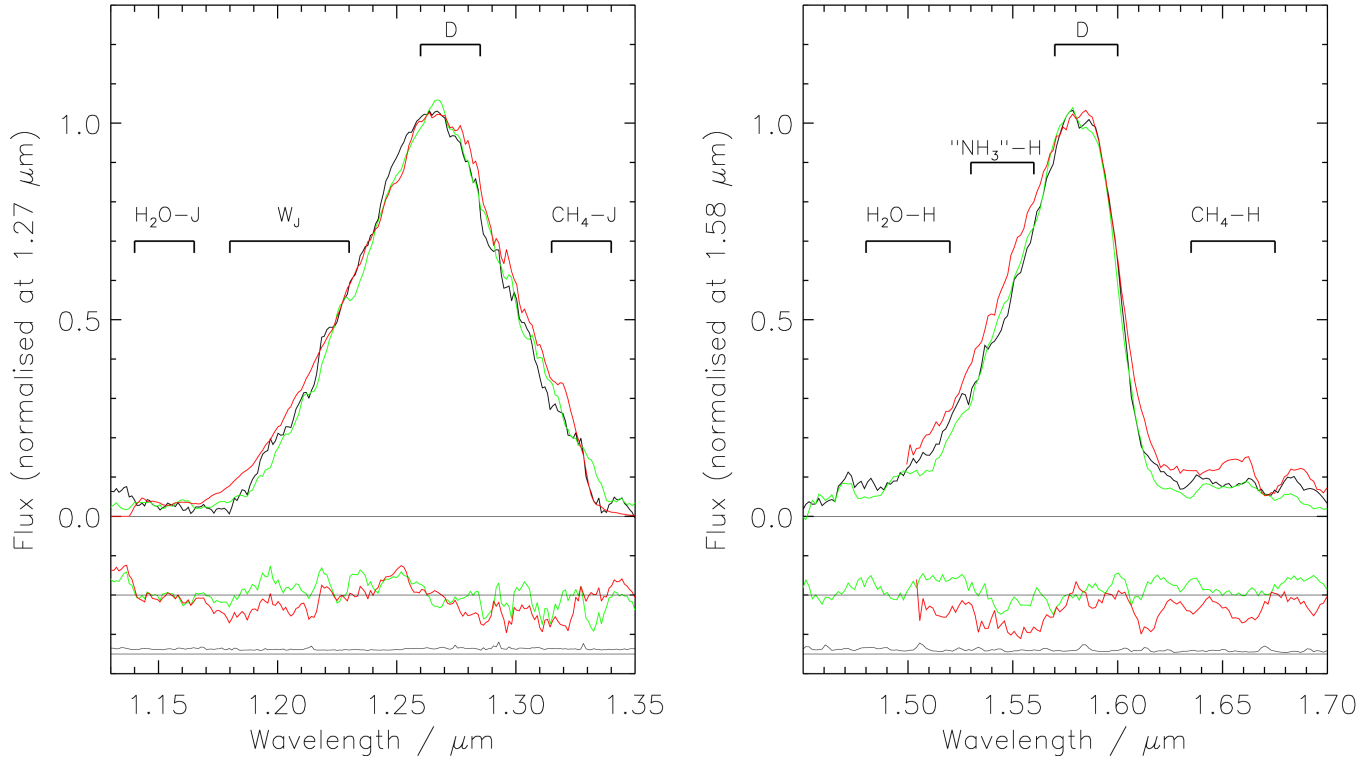}
\caption{The $J$-and $H$-band spectra of ULAS~2146 (black line)
  compared to those of the T8 and T9 spectral standards 2MASS~0415 (red
  line) and ULAS~1335 (green line) respectively \citep{ben08}. The
  numerators for the flux ratios given in Table~\ref{tab:indices} are
  indicated, with the denominators marked with a ``D''.The standard
  spectra have been resampled to the same scales as ULAS~2146, and the
  spectra have been smoothed with a smoothing length of 5 pixels. The
  $J$ and $H$ band spectra for 2MASS~0415 have been taken from
  \citet{mclean03}.
  The single black line in the lowest panel indicates the uncertainty
  spectrum for ULAS~2146. The red and green lines in the middle panel
  indicate the residuals between ULAS~2146 and the spectra of
  2MASS~0415 and ULAS~1335 respectively.
}
\label{fig:compspec}
\end{figure*}

To derive a spectral type for ULAS~2146 we follow the method outlined in
\citet{ben08} for very late T~dwarfs. 
Figure~\ref{fig:compspec} shows the normalised $J$-and
$H$-band\footnote{Normalised to unity at 1.27$\mu$m and 1.58$\mu$m
  respectively} spectra of ULAS~2146 compared to those for the T8 and
T9 spectral templates \citep{burgasser06,ben08}. 
It can be seen from the trace of the residuals between the template
spectra and those of ULAS~2146 that the latter appears to be
intermediate between the two spectral types.

\begin{figure*}
\includegraphics[height=400pt, angle=90]{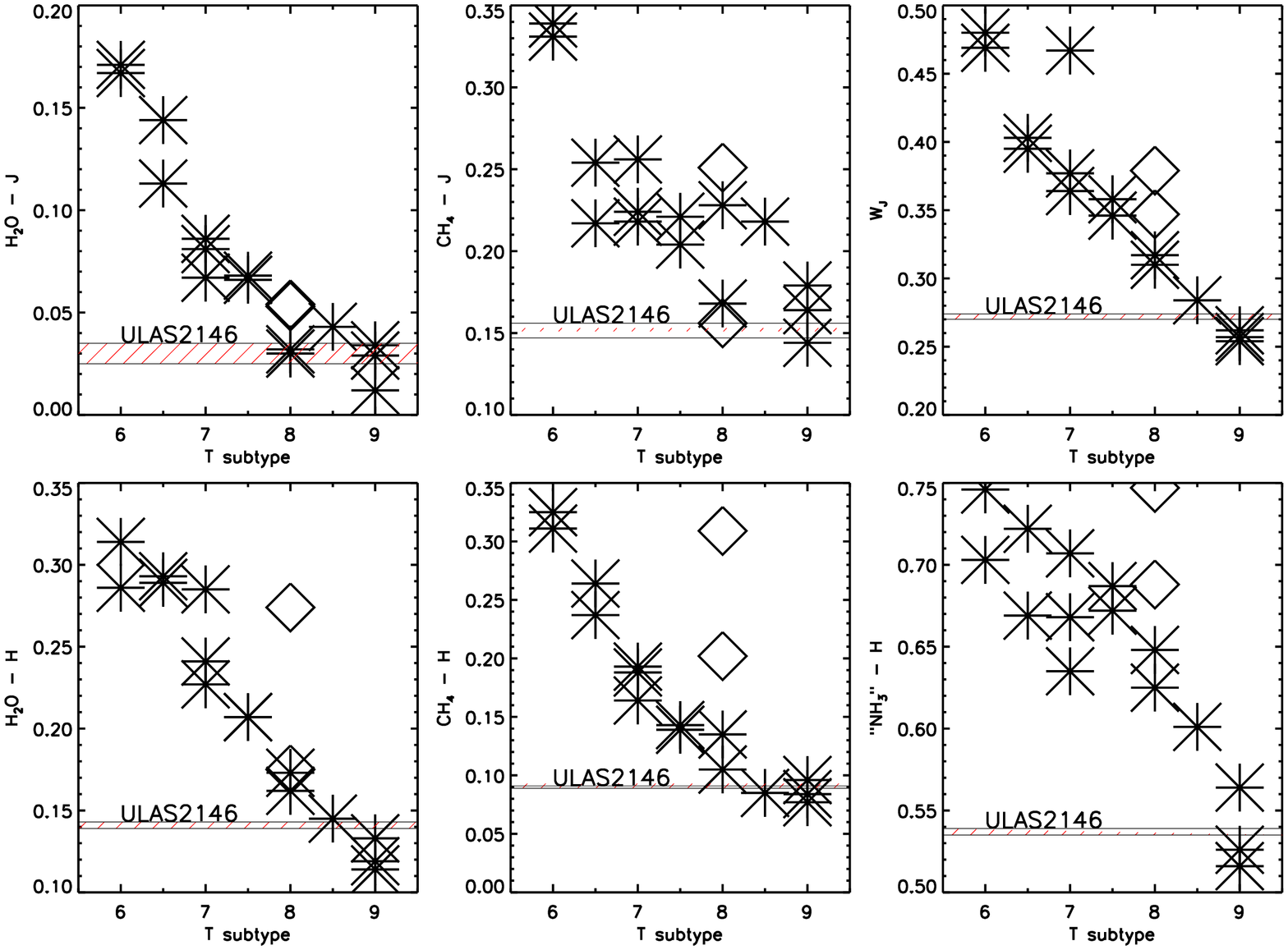}
\caption{Spectral index versus T-subtype for T6-T9 dwarfs. Asterisks
  indicate ``normal'' dwarfs, whilst the diamonds indicate two T8p
  dwarfs: ULAS1017 and 2MASS~J07290002-3954043.
Index values for these objects are drawn from
  \citet{burgasser06,warren07,delorme08,looper07,ben08} or are
  calculated from the objects' spectra supplied by these authors. 
Uncertainties in the indices are smaller than the symbol sizes, whilst
  the uncertainties in the spectral types are typically $<0.5$~subtypes.
The index values for ULAS~2146 are indicated with hatched horizontal
  regions whose height is indicative of the uncertainties.
}
\label{fig:indices}
\end{figure*}

The T~dwarf spectral type indices for ULAS~2146 are given in
Table~\ref{tab:indices}.  
In Figure~\ref{fig:indices} we reproduce Figure~7 from \citet{ben08},
with the spectral type indices for ULAS~2146 indicated along with
those of previously published T6-T9 dwarfs.
As discussed in \citet{ben08}, the H$_2$O-J, CH$_4$-J and CH$_4$-$H$
indices are largely degenerate with type for T8 and T9 dwarfs, whilst
the NH$_3$-$H$ is not yet well understood.
As such, we base our classification on the H$_2$O-$H$ and $W_J$ indices.
It can be seen that the values for its indices are consistent with
classification between T8 and T9, we assign it the type T8.5 ($\pm
0.5$ subtypes).

\begin{table}\renewcommand{\arraystretch}{3}\addtolength{\tabcolsep}{-1pt}
%\centering
\begin{tabular}{c c c c }
  \hline
 {\bf Index} & {\bf Ratio} & {\bf Value} & {\bf Type} \\
\hline
H$_2$O-J & $\frac{\int^{1.165}_{1.14} f(\lambda)d\lambda}{\int^{1.285}_{1.26}f(\lambda)d\lambda }$ & $0.030 \pm 0.005$  & $\geq$ T8 \\[+1mm]
CH$_4$-J & $\frac{\int^{1.34}_{1.315} f(\lambda)d\lambda}{\int^{1.285}_{1.26}f(\lambda)d\lambda }$ &  $0.152 \pm 0.005$ & $\geq$ T8 \\
$W_J$ & $\frac{\int^{1.23}_{1.18} f(\lambda)d\lambda}{2\int^{1.285}_{1.26}f(\lambda)d\lambda }$   &  $0.272 \pm 0.002$  & T9 \\
H$_2$O-$H$ & $\frac{\int^{1.52}_{1.48} f(\lambda)d\lambda}{\int^{1.60}_{1.56}f(\lambda)d\lambda }$ & $0.141 \pm 0.002$  & T8/T9\\
CH$_4$-$H$ & $\frac{\int^{1.675}_{1.635} f(\lambda)d\lambda}{\int^{1.60}_{1.56}f(\lambda)d\lambda }$ & $0.091 \pm 0.002$ & $\geq$ T8 \\
NH$_3$-$H$ &  $\frac{\int^{1.56}_{1.53} f(\lambda)d\lambda}{\int^{1.60}_{1.57}f(\lambda)d\lambda }$ & $0.537 \pm 0.002$ & ... \\
CH$_4$-K &  $\frac{\int^{2.255}_{2.215} f(\lambda)d\lambda}{\int^{2.12}_{2.08}f(\lambda)d\lambda }$ & $0.073 \pm 0.013$ & ... \\
\hline
\end{tabular} 
\caption{The spectral flux ratios for ULAS~2146. Those used for spectral typing are indicated on Figure~\ref{fig:compspec}.}
\label{tab:indices}
\end{table}

\subsection{Keck Laser Guide Star Adaptive Optics Imaging}
\label{subsec:lgs}
To search for possible unresolved binarity, we imaged ULAS~2146 on
03~November~2008~UT using the laser guide star adaptive optics (LGS AO) system
\citep{kecklgsover,kecklgsperf} of the 10-meter Keck
II Telescope on Mauna Kea, Hawaii.  Conditions were photometric with
average seeing.  We used the facility IR camera NIRC2 with its wide
field-of-view camera, which produces an image scale of
$39.69\pm0.05$~mas/pixel.  The LGS provided the wavefront reference
source for AO correction, with the exception of tip-tilt motion.
Tip-tilt aberrations and quasi-static changes in the image of the LGS
as seen by the wavefront sensor were measured contemporaneously with a
second, lower-bandwidth wavefront sensor monitoring the $R=11.5$~mag
nearby star Wolf~940, located 32\arcsec\ away from ULAS~2146. The sodium
laser beam was pointed at the center of the NIRC2 field-of-view for
all observations.

We obtained a series of dithered images, offsetting the telescope by a
few arcseconds, with a total integration time of 360~seconds.  We used
the $CH4s$ filter, which has a central wavelength of 1.592~$\mu$m\
and a width of 0.126~$\mu$m.  This filter is positioned around the
$H$-band flux peak emitted by late-T~dwarfs \citep[see ][]{tinney2005}.  
The images were reduced
in a standard fashion.  We constructed flat fields from the
differences of images of the telescope dome interior with and without
continuum lamp illumination.  Then we created a master sky frame from
the median average of the bias-subtracted, flat-fielded images and
subtracted it from the individual images.  Images were registered and
stacked to form a final mosaic, with a full-width at half-maximum of
0.10\arcsec.  No companions were detected in a $5\arcsec
\times 5\arcsec$ region centered on ULAS~2146.

We determined upper limits on the brightness of potential companions
from the direct imaging by first convolving 
the final mosaic with an analytical representation of the PSF's radial
profile, modeled as the sum of multiple elliptical gaussians.  We then measured
the standard deviation in concentric annuli centered on the science
target, normalized by the peak flux of the targets, and adopted
10$\sigma$ as the flux ratio limits for any companions.  These limits
were verified with implantation of fake companions into the image
using translated and scaled versions of the science target.  

\begin{figure}
\includegraphics[width=200pt,angle=90]{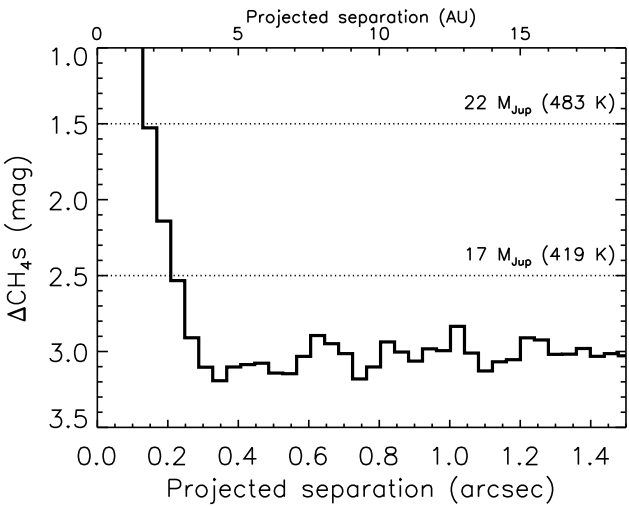}
\caption{Limits on multiplicity of ULAS~2146 based on our Keck LGS
  AO imaging with the $CH4s$ (1.59~$\mu$m) filter and theoretical
  models of \citet{baraffe03}. The masses and $T_{\rm eff}$
  corresponding to hypothetical companions are shown by the dotted
  horizontal lines, assuming an age of 5~Gyr.}
\label{fig:lgslimits}
\end{figure}

Figure~\ref{fig:lgslimits} presents the final upper limits on any
companions.  We employed the COND models of
\cite{baraffe03} to convert the limits into companion
masses, for an assumed age of 5~Gyr and a distance estimate of
12.5~pc (see Section~\ref{sec:binary}).  We assumed any cooler
companions would have similar $(CH4s-H)$ colors to ULAS~2146.

\subsection{Proper motion}
\label{subsec:propermotion}

The photometric follow-up observations that were carried out for ULAS~2146 
provided a second and third epoch of imaging data, showing the
position of the source 0.72 years and 0.93 years after the original
LAS image was measured. 
Although the second epoch  
data covers only a relatively small area of sky (the UFTI mosaic is
135\arcsec\ on each side), we were able to accurately measure the
positions of eight reference stars spread throughout the image, and
match these to their counterparts measured in the LAS images.
We used the same set of reference stars for the LIRIS data.

We used the {\scriptsize IRAF} task {\scriptsize GEOMAP} to derive
spatial transformations from the UFTI and LIRIS $J$-band images into
the LAS $J$-band image in which ULAS~2146 is well detected. 
The transform allowed for linear shifts and rotation, although the
rotation that was required was negligible.
We then transformed the UFTI and LIRIS pixel coordinates of ULAS~2146
into the LAS images using {\scriptsize GEOXYTRAN}, and calculated its change in
position (relative to the reference stars) between the epochs. 
The root-mean-square (rms) scatter in
the difference between the transformed positions of the reference
stars and their actual measured positions was $\sim\pm$0.2 pixels for
the UFTI data and $\sim\pm$0.3 pixels for the LIRIS data
(corresponding to 0.04\arcsec\ and 0.06\arcsec\ in the $J$-band LAS
image).
Assuming that the cardinal axes of the UKIDSS LAS image are aligned
perfectly with the celestial $\alpha$,$\delta$ axes, we thus
determined the proper motion (neglecting parallax) to be $\mu_{\alpha
  cos\delta}=895\pm72$mas/yr, $\mu_{\delta}=-538\pm72$mas/yr.
These uncertainties likely represent an under-estimate, since they do
not include any systematic effects that may be present in data such as
these that were not initially optimised for astrometric use.

As described in Section~\ref{sec:binary}, ULAS~2146 appears to be a
common proper motion companion to the M4 dwarf Wolf~940, which lies at
a distance of 12.50$^{+0.75}_{-0.67}$~pc.
We have thus repeated our proper motion determination incorporating
into the solution the effect of the $79.8 \pm 4.5$ mas parallax
measured for Wolf~940 \citep{USNOpi80}.
Our revised proper motion estimate for ULAS~2146 is thus  $\mu_{\alpha
  cos\delta}=771\pm82$mas/yr, $\mu_{\delta}=-585\pm82$mas/yr, which
incorporates additional uncertainty introduced by the measured
parallax. 
For reasons discussed in Section~\ref{sec:binary}, we consider this
latter proper motion estimate our final value (see also
Table~\ref{tab:2146props}).

\section{A wide binary system}
\label{sec:binary}
A visual comparison between the LAS imaging data and older Schmidt
plate images of the  region around ULAS~2146 revealed the presence of
a high proper motion star just 32\arcsec\ away from the T dwarf. 
This neighbouring source was identified (using the Simbad Database at
CDS) as Wolf~940, a nearby (12.5pc) M4 dwarf with a total proper
motion of 970 mas/yr. The properties of this dwarf are given in
Table~\ref{tab:3708props}, and it can be seen that Wolf~940 and
ULAS~2146 have proper motions that agree to within 1.0$\sigma$.
In order to establish if this pair are a genuine physical binary 
system (as might be inferred from their common proper motion), we have
calculated the expected number of high proper motion stars that might
masquerade as common proper motion companions to LAS very late T dwarf
discoveries. 

Including ULAS~2146, there are now five T8+ dwarfs known with $T_{\rm
  eff}$ estimates ranging down to  $\sim$550--600 K. 
Only ULAS~2146 is known to have a common proper motion companion. 
In general these objects all have $J\simeq$18, and we can thus consider 
ULAS~2146 as a typical example. Distance constraints can be
  estimated based on  spectral type and magnitude, ignoring at this
  stage the association with Wolf~940. A  typical T8 dwarf has
  $M_J$=16.26$\pm$0.37 (Liu et al. 2006) and $T_{\rm eff}\simeq$750 K
  \citep[e.g.][]{saumon07}, which provides a useful upper limit for a
  dwarf with T8.5$\pm$0.5  spectral type. Models
  \citep[e.g.][]{baraffe03} suggest that 750-550 K objects could  have
  $M_J\sim$19 (dependent on age, mass and radius), and for
  $M_J$=15.89--19 the  distance constraint for ULAS~2146
  ($J$=18.21) is 7-29pc. However, we also allow for  the possibility
  that such late T dwarfs could be unresolved binaries, and thus
  potentially  0.75 magnitudes brighter than a single T dwarf. If
  M$_J$=15.14--19, one obtains a more  conservative distance constraint
  of 7-41pc. Note that the known distance of Wolf~940 lies within
  these broad distance constraints, and thus the observable properties
  of ULAS~2146 (spectral type and brightness) are consistent with
  companionship. 

Separations out to $\sim$1 arcminute from the new T8+ population and a
distance range of 7-41pc, corresponds to a space volume of only 0.03
pc$^3$. The local luminosity function measured out to distances of
between 8 and 25 pc \citep[e.g.][]{reid07} suggests space densities of
0.06--0.11 stellar systems pc$^{-3}$, and thus 0.0018--0.0033 stars
actually contained within this volume. 
However, the likelihood of finding a star that had a common proper
motion decreases this number still further. 
We examined all Hipparcos stars from 7-41pc in the direction of
ULAS~2146 ($\pm$45 degs in R.A. and Dec) and found that only three out
of 619 were contained within a 200 mas/yr ($\sim2\sigma$)error circle
centered on the proper motion of the T dwarf. 
This represents a 0.48$\pm$0.28\% probability of finding a
 common proper motion source by chance in such a volume.
We therefore conclude
that we would expect (12$\pm$8)$\times10^{-6}$ stars to masquerade as
common proper motion companions to the T8+ dwarfs discovered in the
UKIDSS LAS to date. 
This represents a vanishingly small probability, and we thus
unambiguously consider that Wolf~940 and ULAS~2146 (here-after
Wolf~940B) are a physical binary system.

\subsection{The properties of Wolf~940A}
\label{subsec:lhs3708a}
This high proper motion M4 dwarf star was first presented by
\citet{wolf1919}, and was first recognised as a high proper motion
object by \citet{rodgers74}. 
Its properties are summarised in Table~\ref{tab:3708props}. 
It has a measured parallax distance of 
12.50$^{+0.75}_{-0.67}$ pc \citep{USNOpi80}, and a mass and metallicity of 
0.27$\pm$0.03 M$_{\odot}$ and -0.06$\pm$0.20 dex respectively, as
estimated from fits  to it's M$_{K_s}$ and V-K colour using the
relations presented in \citet{bonfils05}. 

The kinematics of Wolf~940 are listed in Table~\ref{tab:3708props},
and appear to be consistent with membership of the old disk population -
kinematically defined to have an eccentricity in the UV plane  $<$0.5
and lie outside of the young disk ellipsoid \citep[where the young disk
ellipsoid is defined  as $-20<U<+50$, $-30<V<0$, $-25<W<+10$, see
][]{eggen69,leggett92}. 
The Besan{\c c}on Galactic population synthesis model 
\citep{robin03} reproduces the stellar content of the Galaxy using
various input physical assumptions and a simulated scenario of formation and
evolution. 
This model has  been tuned by comparison against relevant observational data as
described in \citet{haywood97}. 
The disk component of the model comprises numerous sub-populations including 
a 3--5 Gyr population with [Fe/H]=-0.07$\pm$0.18 dex. 
The metallicity and kinematic constraints for Wolf~940A are thus
consistent with an age in the region of 3--5 Gyr.
However, since such kinematic and
compositional arguments apply only to populations of objects, this
line of argument falls short of effectively constraining the age of Wolf~940A.

The study of M dwarf activity by \citet{gizis02} revealed that Wolf~940 has 
H$\alpha$ in absorption, with an equivalent width of 0.262\AA. 
\citet{west08} more recently demonstrated that the drop in activity
fraction (as traced by H$\alpha$) as a function of the vertical
distance from the Galactic plane can be explained by a combination of
thin-disk dynamical heating and a rapid decrease in magnetic activity. 
The timescale for this rapid activity decrease changes according to
the spectral type, and they calibrate this via model fits to a
population of 38,000 SDSS M dwarfs.  
For M4 dwarfs the activity life-time is determined to be
4.5$^{+0.5}_{-1.0}$ Gyr, and we are thus able to put a lower limit of
3.5 Gyr on the age of Wolf~940A from its lack of H$\alpha$
emission.

This limit comes with the caveat that the activity
life-times were derived for a bulk population, and it is not possible
to rule out variability in individual stars. 
Additionally, uncertainty in the spectral type contributes another source of
uncertainty. 
Although the spectral type of Wolf~940A appears to be reliably
determined, we will
consider a worst case scenario to examine the impact this may have on
the age limit. 
For example, were the spectral type of Wolf~940A in error by a whole
subtype, and it was actually an M5 dwarf, we would find  a lower limit
on the age of 6.5~Gyr.  Alternatively, were Wolf~940A an M3 dwarf, the
lower age limit would be 1.5~Gyr. 
Since the uncertainty in the spectral type is certainly much less
than 1 subtype \citep{hawley97}, we adopt the limit
implied by a spectral type of M4.

M~dwarf atmospheres are too cool to produce
H$\alpha$ absorption in the photosphere \citep{cm79,pc81}, and the presence
of H$\alpha$ absorption thus implies the presence of a hot chromosphere 
\citep[i.e. magnetic activity, ][]{cm85,wh08}. 
Therefore, the presence of H$\alpha$
absorption in Wolf~940 indicates that it is still active at some level.  
Indeed, the relative numbers of H$\alpha$ active and inactive M4
dwarfs in SDSS suggests an M4 age no more than $\sim$6 Gyr. The
activity age estimate for Wolf~940 is thus 3.5-6 Gyr, consistent
with the indications from kinematics and metallicity.

\begin{table}
\begin{tabular}{|l|l|}
\hline
Wolf~940A & \\
\hline
R.A. (ep=2000 eq=2000) &  21 46 40.47 \\
Dec (ep=2000 eq=2000)  & -00 10 25.4  \\
R.A. (ep=2007.78 eq=2000) &  21 46 40.89 $^{a}$ \\
Dec (ep=2007.78 eq=2000)  & -00 10 29.5 $^{a}$  \\
PM$_{\alpha \cos{\delta}}$ & $765\pm 2$ mas/yr $^{b}$ \\
PM$_{\delta}$ & $-497\pm 2$ mas/yr $^{b}$ \\
Spectral type & M4 $^{c}$ \\
$V$ & 12.70 $^{c}$ \\
$B-V$ & 1.61 $^{c}$ \\
$J$ & 8.36$\pm$0.02 $^{d}$ \\
$J-H$ & 0.53$\pm$0.04 $^{d}$ \\
$H-K_s$ & 0.34$\pm$0.04 $^{d}$ \\
$V-K_s$ & 5.21 \\
$\pi$ & 79.8$\pm$4.5 mas $^{c}$ \\
Distance & 12.50$^{+0.75}_{-0.67}$ pc \\
m-M & 0.49$\pm$0.13 \\
M$_{K_s}$ & 7.00$\pm$0.13\\
V$rad$ & $-31.6\pm12.2$ km/s $^{e}$ \\
U & $ 34.9\pm6.1$ km/s $^{e}$ \\
V & $-49.4\pm6.0$ km/s $^{e}$ \\
W & $-25.6\pm9.0$ km/s $^{e}$ \\
Galactic speed & 185.3$\pm$8.7 km/s $^{e}$ \\
$[{\rm Fe/H}]$ & -0.06$\pm$0.20 $^{f}$ \\
Mass & 0.27$\pm$0.03 M$_{\odot}$ $^{f}$ \\
H$_{\alpha}$EW & 0.262 \AA$^{g}$ \\
Age & 3.5--6.0 Gyr $^{h}$ \\
\hline
\multicolumn{2}{|l|}{$^a$ Epoch of the UKIDSS LAS observation}\\
\multicolumn{2}{|l|}{$^b$ \citet{USNOpi80}}\\
\multicolumn{2}{|l|}{$^c$ \citet{reid95}}\\
\multicolumn{2}{|l|}{$^d$ From 2MASS database}\\
\multicolumn{2}{|l|}{$^e$ \citet{dawson2005}}\\
\multicolumn{2}{|l|}{$^f$ Based on polynomial relationships (functions of}\\
\multicolumn{2}{|l|}{V-K$_s$ and M$_{K_s}$) from \citet{bonfils05}}\\
\multicolumn{2}{|l|}{$^g$ \citet{gizis02}}\\
\multicolumn{2}{|l|}{$^h$ Derived from activity life-time information presented}\\
\multicolumn{2}{|l|}{in \citet{west08}}
\end{tabular}
\caption{Properties of Wolf~940A.}
\label{tab:3708props}
\end{table}

\subsection{The properties of Wolf~940B (ULAS~2146)}
\label{subsec:lhs3708b}

As a companion, Wolf~940B will share the same distance as the
M4 primary  (12.50$^{+0.75}_{-0.67}$ pc), and the two objects thus have
a projected line-of-sight separation of 400AU. 
The actual
semimajor axis of the binary depends on its orbital parameters.
Following the method of \citet{torres99}, we assume random
viewing angles and a uniform eccentricity distribution between
0~$<~e~<$~1 to derive a correction factor of 1.10$^{+0.91}_{-0.36}$
(68.3\% confidence limits) to convert projected separation into
semimajor axis.  At a distance of 12.50$^{+0.75}_{-0.67}$~pc, this
results in a semimajor axis of 440$^{+370}_{-150}$~AU.  For a total
mass of 0.27~\Msun\ (i.e. neglecting the mass of Wolf 940B), this
corresponds to an orbital period of 18000$^{+26000}_{-8000}$~years.
This is quite typical
when compared to other brown dwarfs in widely separated binary systems
\citep[e.g. ][]{burgasser05,pinfield06}.

It has been known for some time that the degree of multiplicity
amongst very young stars is greater than that of the more evolved
field star populations \citep{duq91,leinert93}, and thus that the
majority of binary systems form together in their nascent
clouds. Binary components can therefore generally be assumed to share
the same age and composition, and we therefore assume that Wolf~940B
also has an age of 3.5-6~Gyr and a composition of $[{\rm
    Fe/H}]$=-0.06$\pm$0.20.

\begin{figure}
\includegraphics[height=250pt, angle=0]{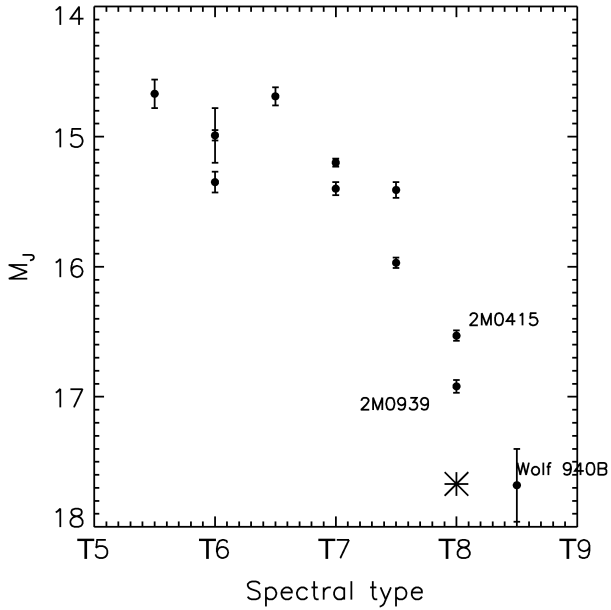}
\caption{M$_J$(MKO) vs spectral type for T~dwarfs later than T5 with
  parallaxes. Values of $M_J$ have been taken from
  \citet{knapp04}, whilst the spectral types are on the system of
  \citet{burgasser06} and \citet{ben08}.The three latest type T dwarfs
  with parallaxes and M$_J$(MKO) are labeled. The M$_J$ inferred for
the components of 2MASS~J0939-2448 \citep[see text and
  ][]{burgasser08a} is indicated with an asterisk.}
\label{fig:mjplot}
\end{figure}

Since we have an accurate distance for Wolf~940B we can determine its
absolute magnitude, and values for $M_J$, $M_H$ and $M_K$ are given in
Table~\ref{tab:2146props}.
Figure~\ref{fig:mjplot} shows M$_J$ plotted against spectral type for this
and other T~dwarfs with reliably determined parallaxes. 
The very faint nature of Wolf~940B is apparent, suggestive of a very
low $T_{\rm eff}$.
2MASS~J0939-2448, which has recently been suggested as an equal mass
binary system with component $T_{eff} \sim 600$K \citep{burgasser08a},
is indicated.
The inferred $M_J$~=~17.67 for the individual components of
2MASS~J0939-2448 is strikingly similar to that observed for Wolf~940B.

To determine the bolometric flux, $F_{bol}$, from Wolf~940B, we have
combined our observed $JHK$ spectra (flux calibrated using our UFTI
follow-up photometry) with model spectra that allow us to estimate the
flux contributions from regions shortward and longward of our
near-infrared spectral coverage. 
We scaled the $\lambda < 1.0 \mu$m portion of the model spectra to
match the short wavelength end of our $J$-band spectrum, whilst we used
our $L'$-band photometry to scale the  $\lambda > 2.4 \mu$m portion of
the model spectra. 
We then joined them to our observed spectra and
estimated the bolometric flux, assuming all flux emerges between 0.5
$\mu$m and 30$\mu$m. 
Provided the measured $L'$-band photometry is in a band where the
level of emitted flux is relatively high (compared to the unmeasured
wavelength regions), and that the  theoretical models can be relied upon to
provide a ``reasonable'' approximation to the  shape of the spectral
energy distribution \citep[c.f. ][]{mainzer07,cushing06,roellig04},
then this approach should provide valid results. 

\begin{figure}
\includegraphics[height=250pt, angle=90]{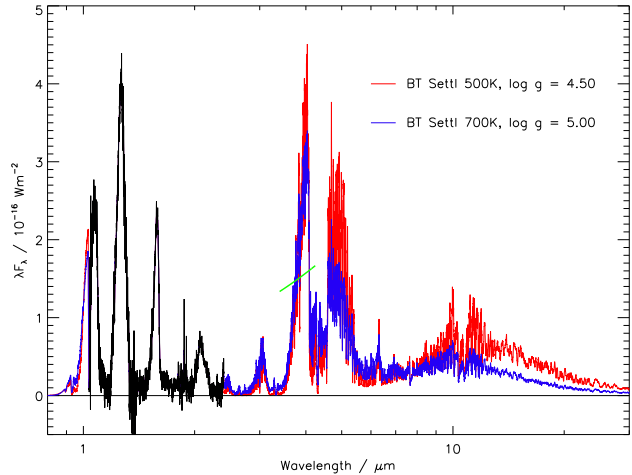}
\caption{The NIRI $JHK$ spectrum of Wolf~940B joined to scaled BT-Settl
model spectra bracketing the range of parameter space considered for
our $F_{bol}$ estimate. The mean flux level implied by the $L'$-band photometry
is indicated with a short green line.}
\label{fig:boloplot}
\end{figure}

We have used the latest BT-Settl solar metallicity model spectra
covering the 500--700K temperature range, with log~{\it g} = 4.5--5.0,
to provide the normalised shorter and longer wavelength spectral
extensions to the observed near-infrared spectrum. 
We then took the median as our final value for $F_{bol}$.
The scatter in values was taken as an estimate of the systematic
uncertainty associated with our use of these normalised theoretical extensions.
Figure~\ref{fig:boloplot} shows the scaled spectra for two extremes of
our theoretical extension parameter space, along with our observed
$JHK$ spectrum for Wolf~940B.
It can be seen that the optical region contributes a very small portion of the
total flux ($\sim$2\%), and its associated uncertainty is thus of only minor
significance.
Longward of the $K$-band, however, the contribution to the overall
flux is greater, with $\sim$60\% of the flux emitted with $\lambda >
2.4 \mu$m.
We found that experimenting 
with the range of normalised theoretical spectra introduced an
uncertainty of $\pm 9$\%. The uncertainties in our photometry
used for scaling the model spectra introduced an additional $\sim 5$\%
uncertainty to our bolometric flux estimate, dominated
by the contribution from our $L'$-band measurement. 
Our total uncertainty in $F_{bol}$ is thus $\pm 10$\%, with a final
value of $1.75 \pm 0.18 \times 10^{-16}$~Wm$^{-2}$.
The luminosity of Wolf~940B then comes directly 
from the bolometric flux and distance, allowing for uncertainties in
both (see Table~\ref{tab:2146props}).

To place constraints on the mass and radius of Wolf~940B we rely on
theoretical structure models and infer these properties from the
luminosity and age information. 
This requires an assumption about the multiplicity of Wolf~940B. 
Section~\ref{subsec:lgs} shows that 
Wolf~940B is unresolved at an angular resolution of 0.1'' corresponding
to a spatial resolution of 1.2 AU at the distance of the system. 
So although there is evidence \citep{burgasser05} that the binary
fraction of brown dwarfs (as resolved at such resolutions) in widely
separated stellar-brown dwarf multiple systems is notably higher
(45$^{+15}_{-13}$\%) than that of field brown dwarfs in analogous
samples (18$^{+7}_{-4}$\%), Wolf~940B is not one of these systems.  
To assess the likelihood that it may be a tighter 
unresolved binary system we considered the analysis of
\citet{maxted05}, who use Monte Carlo simulation techniques to assess
radial velocity survey data and find that a Gaussian separation
distribution with a peak at 4AU and a standard deviation
$\sigma_{log(a/au)}=0.6-1.0$ correctly predict the number of observed
binaries (radial velocity variables). 
Their estimated total binary fraction is 32-45\%, consistent with
estimates from open cluster studies \citep[e.g.][]{lod07a,pinfield03}.  
For a Gaussian separation distribution of this type, we would expect
at most a 10-20\% binary fraction for systems with separation
$<$1.2AU.

The metastudy by \citet{allen07} estimates the binary fraction for
objects later than M6 in the field as 20--22\%, using bayesian
methods, and $\sim$6\% for systems with seperations less than 1~AU.
Similarly, \citet{joergens08} find a binary fraction for low-mass
stars and brown dwarfs of 10--30\% in the Chameleon I star forming
region, with a frequency of less than 10\% for binary separations of $<$1~AU.

Since the binary fraction of a magnitude limited survey, such as
UKIDSS, will be increased by unresolved binaries, which are seen to greater
distances, caution must be used when assessing the likelihood of
binarity for an object from such a survey. 
For this reason we do not derive a formal likelihood of
unresolved binarity for Wolf~940B.
However, based on the results described above, it is possible to state
that it is likely that Wolf~940B is a single object, and we  proceed with our
analysis on this basis.

Assuming  evolution of the mass-luminosity and radius-luminosity relations from
\citet{baraffe03} isochrones, we used linear interpolation (between
isochrones) to derive theoretical mass and radius estimates
appropriate for the measured luminosity and age constraints of
Wolf~940B. 
These parameters (including $\log{g}$) are given in
Table~\ref{tab:2146props}.

We can use our estimate of the radius and the luminosity to determine
 $T_{\rm eff}$. 
However, since the radius estimate depends strongly on the assumed
 age, so does the derived  $T_{\rm eff}$.  
As discussed in Section~\ref{subsec:lhs3708a}, the spectral type that
is used for Wolf~940A influences the age constraints implied by its
H$\alpha$ absorption. The worst case scenarios of errors of $\pm 1$ subtype in
spectral type would lead to age ranges of 1.5--10 Gyr, or
6.5--10~Gyr.
The extremes of these alternatives would imply radii of $0.105
R_{\odot}$ for an age of 1.5~Gyr and $0.084 R_{\odot}$ for an age of
10~Gyr, and $\log{g}$ constraints of 4.50---5.1 respectively.
Given our luminosity estimate, these extreme cases imply $T_{\rm eff}$=540~K
and $T_{\rm eff}$=605~K respectively.
Since the spectral type uncertainty is small, however, we adopt the
range of values implied by the M4 classification, and the associated
 age estimate of 3.5--6~Gyr, in Table~\ref{tab:2146props}.

We thus obtain our best estimate of $T_{\rm eff}$=570$\pm$25~K and
$\log{g}$ = 4.75--5.00 for Wolf~940B directly from the constraints we
place on the luminosity and radius of this object.

\begin{table}
\begin{tabular}{|l|l|}
\hline
Wolf~940B (ULAS~2146) & \\
\hline
R.A. (ep=2000 eq=2000) &  21 46 38.41 \\
Dec (ep=2000 eq=2000)  & -00 10 34.6 \\
R.A. (ep=2007.78 eq=2000) &  21 46 38.83 $^{a}$ \\
Dec (ep=2007.78 eq=2000)  & -00 10 38.7 $^{a}$  \\
$\mu_{\alpha \cos{\delta}}$ & $771\pm82$ mas/yr \\
$\mu_{\delta}$ & $-585\pm82$ mas/yr \\
Spectral type & T8.5$\pm$0.5 \\
Separation & 32\arcsec\ \\
           & 400$\pm$22 AU $^{b}$ \\
$J$ & 18.16$\pm$0.02 \\
$z_{\sc EFOSC2}-J$ & 3.99 $\pm$ 0.13 \\
$Y-J$ & 0.81$\pm$0.04 \\
$J-H$ & -0.61$\pm$0.04 \\
$H-K$ & -0.08$\pm$0.05 \\
$J-K$ & -0.69$\pm$0.05 \\
$L'$ & $15.38 \pm 0.1$ \\
$K-L$ & $3.47 \pm 0.11$ \\
Bolometric flux & $1.75 \pm 0.18  \times 10^{-16}$Wm$^{-2}$ $^{c}$ \\
$M_J$ & 17.68$\pm$0.28 $^{b}$ \\
$M_H$ & 18.29$\pm$0.28 $^{b}$ \\
$M_K$ & 18.37$\pm$0.28 $^{b}$ \\
$\log (L/L_{\odot})$ & $-6.07 \pm 0.04^{b}$ \\
$[{\rm Fe/H}]$ & $-0.06\pm0.20$ $^{d}$ \\
Mass & 20--32 M$_{J}$ $^{e}$ \\
Radius & $0.094 \pm 0.004 R_{\odot}$ $^{e}$ \\
$\log~g$ & 4.75--5.00 $^{e}$ \\
$T_{\rm eff}$ & $570 \pm 25$K $^{f}$ \\
\hline
\multicolumn{2}{|l|}{$^a$ Epoch of the UKIDSS LAS observation}\\
\multicolumn{2}{|l|}{$^b$ Inferring a distance of
  12.50$^{+0.75}_{-0.67}$ from Wolf 940A}\\
\multicolumn{2}{|l|}{$^c$ Integrating the measured flux from 1.0--2.4 microns}\\
\multicolumn{2}{|l|}{and adding a theoretical correction at longer and}\\
\multicolumn{2}{|l|}{shorter wavelength (see text).}\\
\multicolumn{2}{|l|}{$^d$ Inferred from Wolf 940A}\\
\multicolumn{2}{|l|}{$^e$ Constraints derived from structure models as a}\\
\multicolumn{2}{|l|}{function of luminosity for ages 3.5--6 Gyr}\\
\multicolumn{2}{|l|}{$^f$ Derived from the luminosity and radius constraints}
\end{tabular}
\caption{Properties of Wolf~940B (ULAS~2146).}
\label{tab:2146props}
\end{table}

\section{Testing the models}
\label{sec:nirspecfit}

\begin{figure}
\includegraphics[height=250pt, angle=90]{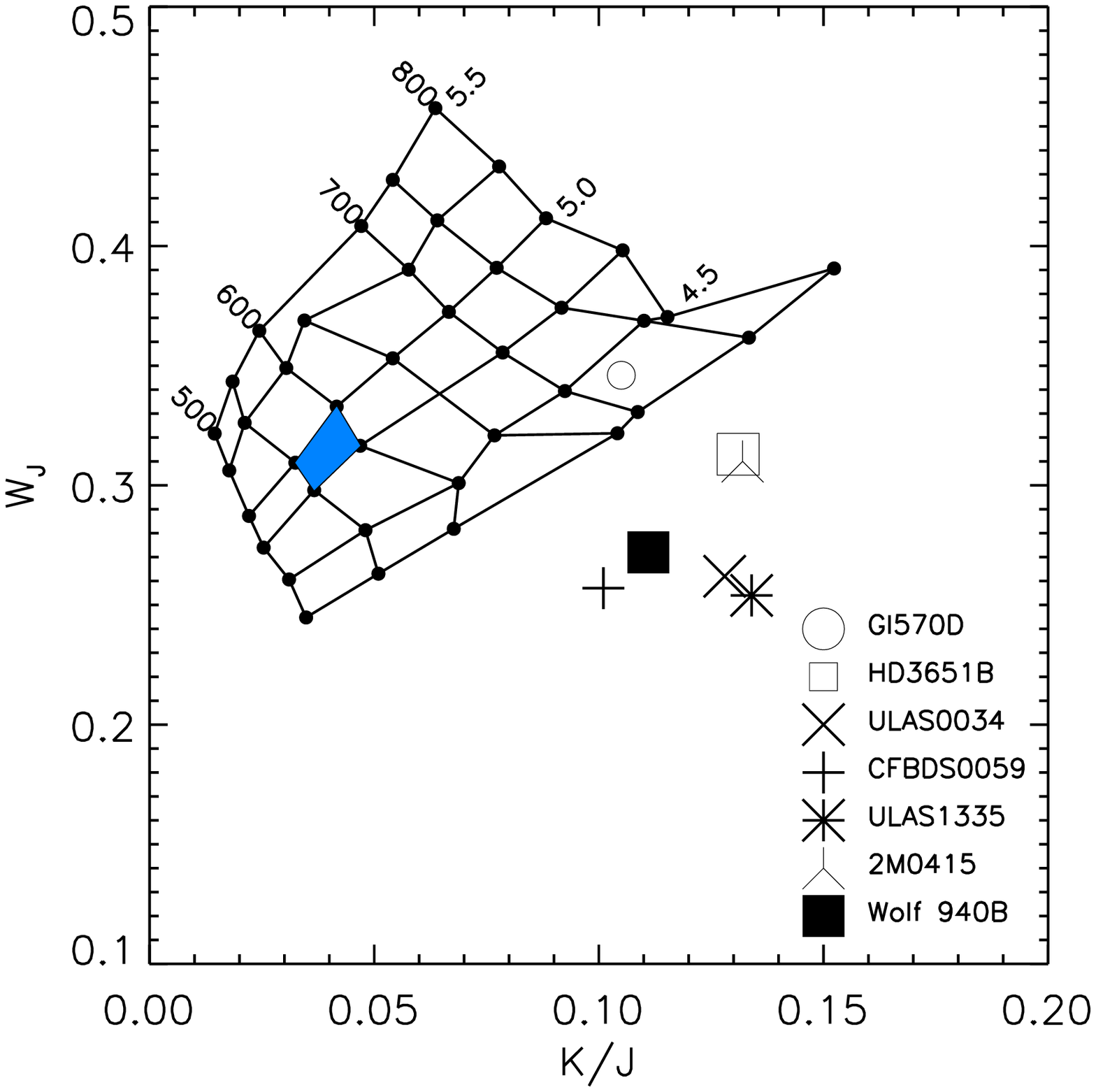}
\caption{$W_J$ versus $K/J$ indices for a grid of the most recent set
  of solar metallicity BT-Settl model spectra. Flux ratios for HD~3651B
  were measured using the spectrum from \citet{burgasser07}, for
  2MASS~0415 using the spectrum from \citet{burgasser07}, and for Gl~570D
  using the spectrum from \citet{geballe01}. Those for
  the three T9 dwarfs ULAS~0034, CFBDS~0059 and ULAS~1335 were taken from
  their respective discovery papers
  \citep{warren07,delorme08,ben08}. The blue shading indicates the
  region of this grid that should contain Wolf~940B. Uncertainties are
  of similar size to the symbols.}
\label{fig:kjwj_noanchor}
\end{figure}

We now use our robust properties for Wolf~940B to make a direct
comparison between observation and specific theoretical model predictions.
We first apply the ($W_J$,$K/J$) analysis described by \citet{warren07} for
ULAS~0034, and since repeated for CFBDS~0059
\citep{delorme08} and ULAS~1335 \citep{ben08}. 
In Figure~\ref{fig:kjwj_noanchor} we plot a grid of the $W_J$ and $K/J$
flux ratios for a recent set of solar metallicity BT-Settl models,
along with the same ratios for a group of late-T~dwarfs, including three
T9s, the benchmark T~dwarfs HD~3651B and Gl~570D, along with the T8
dwarf 2MASS~0415. 
It is immediately clear that there are large differences between the
model predictions and the values derived from the objects' spectra.

\begin{figure}
\includegraphics[height=250pt, angle=90]{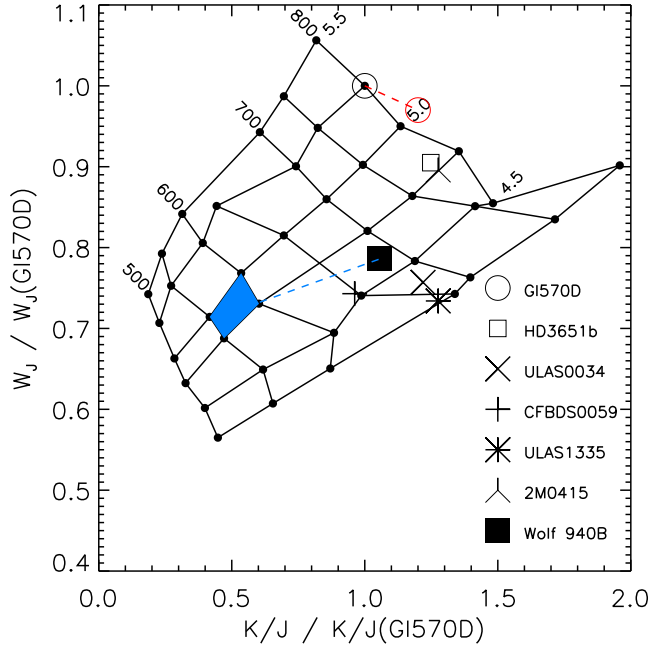}
\caption{$W_J$ versus $K/J$ indices for a grid of solar metallicity
  BT-Settl model spectra. The grid is normalised such that the model
  values for the $T_{\rm eff}$ and log~{\it g} of Gl~570D lie at
  coordinates (1,1). All observed values of $W_J$ and $K/J$ are shown
  as a proportion of those of Gl~570D.  We have indicated, with a
  red dotted line and open circle, the
  relative shift in position on the grid associated with increasing
  metallicity by +0.1 dex. Uncertainties are
  of similar size to the symbols.The blue dotted line highlights the
  difference between the true properties of Wolf 940B, and those
  expected from the model grid.}
\label{fig:kjwj}
\end{figure}

To assess the ability of the model spectra to make relative
predictions for objects' properties, we replot the same grid in
Figure~\ref{fig:kjwj}, however in this case we anchor the grid to 
values for the benchmark T~dwarf Gl~570D.
We have adopted the values derived by \citet{saumon06} of $T_{\rm
  eff} = 810 \pm 10$K and $\log g = 5.09 - 5.23$, and use the metallicity
found by \citet{geballe01} of [M/H] $= 0.01$. 
For the purposes of anchoring the solar metallicity ($W_J$,$K/J$) grid
we associated Gl~570D with the model values for $T_{\rm eff} = 800$K,
and $\log g = 5.25$.  
Even with such correction this diagram still fails to correctly
identify the properties of Wolf~940B.
The model spectra under-predict the $K/J$ ratio in absolute
terms (Figure~\ref{fig:kjwj_noanchor}), and also predict a greater decrease in
its value on going from $T_{\rm eff} = 800$K to $T_{\rm eff} = 570$K
than is observed from Gl570D and Wolf~940B (Figure~\ref{fig:kjwj}).
In the case of the $W_J$ value the absolute prediction is an
over-estimate, whilst the predicted decrease between $T_{\rm eff} =
800$K and $T_{\rm eff} = 570$K is close to reality.
As a result of these effects, in the case of
Wolf~940B, a simple ($W_J$,$K/J$) analysis would have over-estimated
the temperature by $\sim 100$K.

\begin{figure}
\includegraphics[height=350pt, angle=0]{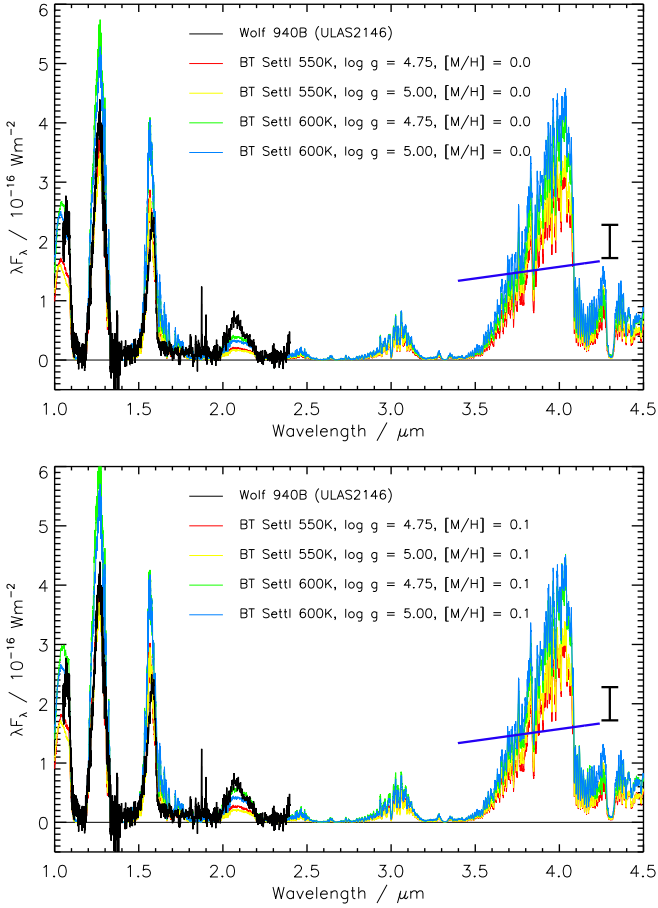}
\caption{A comparison of BT-Settl model spectra bracketing the derived
  parameters for Wolf~940B with the observed spectrum. The top plot
  shows the comparison for solar metallicity models, whilst the bottom
  plot show mildly metal rich models.The short, straight, blue line in
  each case indicates the mean flux level in our $L'$-band photometric
  observation.The black error bar at the right of each plot is
  representative of the uncertainty in a scaled model flux of $2
  \times 10^-16$W$m^{-2}$ due to the uncertainties in the parallax and
  radius of Wolf~940B.}
\label{fig:spec_comp}
\end{figure}

\begin{figure}
\includegraphics[height=350pt, angle=0]{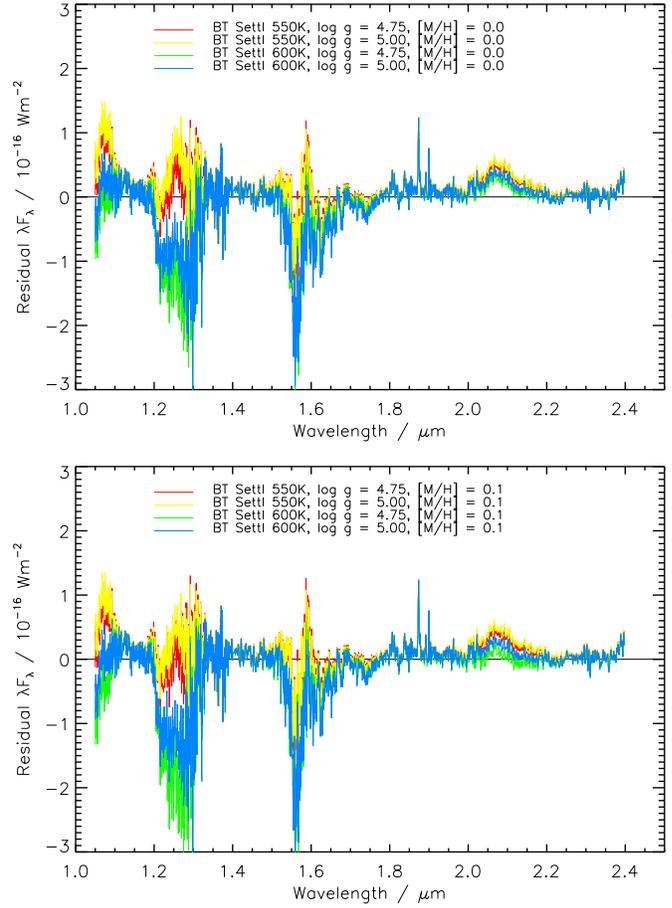}
\caption{The residuals between the model spectra plotted in
  Figure~\ref{fig:spec_comp} and the $JHK$ spectrum of Wolf~940B.}
\label{fig:spec_resid}
\end{figure}

In Figure~\ref{fig:spec_comp} we plot the comparison of the observed
spectrum of Wolf~940B with those of the models that bracket its derived
properties, scaled for the distance of 12.5pc and a radius
0.094~\Rsun, and in Figure~\ref{fig:spec_resid} we plot the residuals
between $JHK$ model spectra and our data.
The reason for the offsets in the ($W_J$,$K/J$) plots is
demonstrated here.
The low predicted value of $K/J$ in both the un-anchored
and the anchored plots appear to be driven by systematic underestimate
of the $K$-band flux suggesting problems with the opacity due to
collisionally induced absorption (CIA) by H$_2$.
The only model that does not underestimate the $K$-band peak is that for
$T_{\rm eff} = 600$K, $\log g = 4.75$, [M/H]~=~+0.1. 
However, this model spectrum overestimates the $J$-and $H$-band peaks
by the greatest extent.
Since model spectra for the low-metallicity
parameter space are yet to be computed, we are unable to explore the full
range of possibilities for this object's metallicity. In fact,
\citet{liu07} have found that the relative changes in model spectra
with metallicity do not agree very well with the available data for
late-T~dwarfs. We thus defer a
more extensive model comparison to a future paper.

Examination of the strengths of the $YJHK$ flux peaks, with reference
to the trends in model spectra
with metallicity, gravity and temperature is thought to be useful for
identifying T~dwarfs with unusual properties \citep[e.g.][]{pinfield08}.
The well constrained nature of Wolf~940B makes it a useful
reference point for assessing such spectral sensitivities.
We have calculated the $JHK$ flux peak ratios for Wolf~940B, and
compare them to other very late-T~dwarfs and the two late-T benchmarks
Gl~570D and HD~3651B in Table~\ref{tab:fpeaks}. 
Flux peak ratios involving the $Y$-band
peak must be neglected, however, since our $JHK$ spectrum for ULAS~2146
does not provide sufficient coverage.

As discussed in more detail by \citet{pinfield08},
the most dramatic trends in the
relative strengths of the flux peaks of the BT-Settl model spectra appear
to be associated with varying gravity and metallicity, with $K$-band
suppression (decreasing $K/J$) 
seen with increasing gravity and decreasing metallicity.
The $H$-band peak is more weakly affected by varying these parameters,
and has trends in the opposite sense to the $K$-band, i.e. decreasing
strength with decreasing gravity or metallicity (decreasing $H/J$).
The flux peaks show a weaker response to varying $T_{\rm eff}$, with
$H$-band strengthening with falling $T_{\rm eff}$, and the $K$-band
weakening.

Direct comparison of the flux peak ratios of Wolf~940B with the other
two benchmark objects in Table~\ref{tab:fpeaks} must be rather cursory
at this stage.
The $\sim 200$K difference in $T_{\rm eff}$ between the objects makes
it impossible to disentangle the effects of gravity/metallicity from
those due to $T_{\rm eff}$ differences. However, it does appear that
the relative ratios of the three objects are broadly consistent with
their properties and the trends described above.

\begin{table}
%\centering
\begin{tabular}{c c c c c}
  \hline
 {\bf Object} & {\bf Sp. Type} &{\bf $H/J$} & {\bf $K/J$} & {\bf $K/H$} \\
\hline
Wolf~940B (ULAS~2146)& T8.5& 0.454 & 0.111 & 0.245 \\
ULAS~0034$^{1}$ &  T9 & 0.475 & 0.126 & 0.266 \\
CFBDS~0059$^{2}$ & T9 & 0.580 & 0.095 & 0.164 \\
ULAS~1335$^{3}$  & T9 & 0.574 & 0.134 & 0.232 \\
2MASS~0415$^{4}$ & T8 & 0.531 & 0.132 & 0.249 \\
2MASS~0939$^{5}$ & T8 & 0.483 & 0.060 & 0.124 \\
2MASS~0729$^{6}$ & T8p & 0.447 & 0.093 & 0.207 \\
ULAS~1017$^{3}$  & T8p & 0.411 & 0.119 & 0.288 \\
Gl~570D$^{7}$    & T7.5 & 0.423 & 0.105 & 0.247 \\
HD~3651B$^{8}$   & T7.5 & 0.454 & 0.131 & 0.289 \\	

\hline
\multicolumn{4}{|l|}{Original publications for source spectra: }\\
\multicolumn{4}{|l|}{$^1$ \citet{warren07}}\\
\multicolumn{4}{|l|}{$^2$ \citet{delorme08}}\\
\multicolumn{4}{|l|}{$^3$ \citet{ben08}}\\
\multicolumn{4}{|l|}{$^4$ \citet{burgasser04}}\\
\multicolumn{4}{|l|}{$^5$ \citet{burgasser06}}\\
\multicolumn{4}{|l|}{$^6$ \citet{looper07}}\\
\multicolumn{4}{|l|}{$^7$ \citet{geballe01}}\\
\multicolumn{4}{|l|}{$^8$ \citet{burgasser07}}\\
\end{tabular} 
\caption{The flux peak ratios for other published T8+
  dwarfs, along with several T8 dwarfs, and the two late-T
  benchmarks. We have included the values for Wolf~940B from
  Table~\ref{tab:indices} for comparison.
The ranges for each flux peak are as follows:
  $J$:~1.25--1.29 $\mu$m; $H$:~1.56--1.60 $\mu$m; $K$:~2.06--2.10
  $\mu$m.}
\label{tab:fpeaks}
\end{table}

\section{Summary and conclusions}
\label{sec:summ}

We have identified a T8.5 ($\pm 0.5$ sub-types) dwarf in a common
proper motion binary system with the M4 dwarf Wolf~940. 
We designate
the M4 primary as Wolf~940A, and the T8.5 dwarf as Wolf~940B, which
lies at a projected separation of 400AU. 
By using the properties of
Wolf~940A to constrain those of Wolf~940B, and with reference to
evolutionary structural models of \citet{baraffe03} we estimate that
Wolf~940B has $T_{\rm eff} = 570 \pm 25$K, $\log g = 4.75-5.0$ and
[M/H]~$= -0.06 \pm 0.20$. This
represents the first estimate of the properties for a T8+ dwarf that
does not rest on the fitting of model spectra, although we do rely on
the radius predicted by evolutionary models and an age estimate from
the activity of the primary of 3.5--6.0 Gyr.

Our comparison of the near-infrared spectrum for Wolf~940B with the
current generation of BT-Settl model spectra reveals that the strength
of the $K$-band flux peak is underestimated by the models.
This is likely
the driving factor behind the $+100$K temperature over-estimate
implied by  ($W_J$,$K/J$) spectral ratio analysis.
This indicates that $T_{\rm eff}$ determined for late-T~dwarfs from
($W_J$,$K/J$) analysis should be treated with extreme caution.

This system should be of significant benefit for improving understanding
of $<600$K atmospheres. 
In the near future we expect Spitzer IRS spectroscopy and IRAC imaging to
be obtained, which will allow a detailed
examination of the predictions of a variety of very cool model
spectra.
Furthermore, intermediate resolution spectroscopy can be used to
assess if Wolf~940B is a rapid rotator, whilst repeat observations can be
used to search for close binarity via modulation of Wolf~940B's radial
velocity.
Finally, it is highly desirable that we improve the metallicity
constraints on this system. 
This requires both improvement in the understanding of metallicity
indicators in M~dwarfs, and a more detailed study of Wolf~940A.
Repeat H$\alpha$ measurements for Wolf~940A will reveal if its
H$\alpha$ absorption is stable, thus providing an indication of the
reliability of our age constraint.

\section*{Acknowledgments}

Gemini Observatory is operated by the Association of Universities for
Research in Astronomy, Inc. (AURA), under a cooperative agreement with
the NSF on behalf of the Gemini partnership: 
the National Science Foundation (United States),
the Science and Technology Facilities Council (United Kingdom), the
National Research Council (Canada), CONICYT (Chile), the Australian
Research Council (Australia), Ministério da Ciência e Tecnologia
(Brazil) and SECYT (Argentina).

SKL is supported by the Gemini Observatory, which is operated by AURA,
on behalf of the international Gemini partnership of Argentina,
Australia, Brazil, Canada, Chile, the United Kingdom, and the United
States of America. 

MCL and TJD acknowledge support for this work from NSF grant
AST-0507833 and an Alfred P. Sloan Research Fellowship.

This research has made use of the SIMBAD database,
operated at CDS, Strasbourg, France, and has benefited from the SpeX
Prism Spectral Libraries, maintained by Adam Burgasser at
http://www.browndwarfs.org/spexprism.

We gratefully acknowledge the Keck LGS AO team for their exceptional
efforts in bringing the LGS AO system to fruition.  It is a pleasure
to thank Randy Campbell, Cynthia Wilburn, and the Keck Observatory
staff for assistance with the observations.
\bibliographystyle{mn2e}
\bibliography{refs}

\end{document}